\documentclass{emulateapj}

\usepackage{subequation}
\journalinfo{ApJ, in press (v. 661, June 2007)}

\def\solphys{{Sol. Phys.,}}

\def\apss{{Astrophys. Space Sci.,}}

\newcommand{\velunits}{~$\rm km~s^{-1}$}

\newcommand{\specangunits}{~$\rm dyne~cm~sr^{-1}$}

\shorttitle{Solar Wind Angular Momentum}
\shortauthors{Li, Habbal, \& Li}

\begin{document}

\title{Angular momentum transport and proton-alpha differential streaming
    in the solar wind}
\author{Bo Li}
\affil{Institute of Mathematical and Physical Sciences, University of Wales Aberystwyth,
  SY23 3BZ, UK}
\email{bbl@aber.ac.uk}
\author{Shadia Rifai Habbal}
\affil{Institute for Astronomy, University of Hawaii, Honolulu, HI 96822, USA}
\and
\author{Xing Li}
\affil{Institute of Mathematical and Physical Sciences, University of Wales Aberystwyth,
  SY23 3BZ, UK}

\begin{abstract}
The effect of solar rotation on the proton-alpha differential flow speed, $v_{\alpha p}$,
       and consequently on the angular momentum transport in the solar wind, is explored.
Using a 3-fluid model, it is found that the radial component of the momentum equation for ions
       is modified by the force introduced by the azimuthal components. 
This force plays an important role in the force balance in interplanetary space, 
       hence impacts the radial flow speed of the species considered, bringing them closer to each other.
For the fast solar wind, the model cannot account for the decrease of $v_{\alpha p}$ observed
       by Helios between $0.3$ and $1$~AU. 
However, it is capable of reproducing the profile of $v_{\alpha p}$ measured by Ulysses beyond 2 AU,
       if the right value for $v_{\alpha p}$ is imposed at that distance. 
In the slow solar wind, on the other hand, the effect of solar rotation seems to be more pronounced
       if one starts with the value measured by Helios at 0.3 AU. 
In this case, solar rotation introduces a relative change of 10-16\% in the radial component of the flow
       speed of the alpha particles between 1 and 4 AU. 
The model calculations also show that, although alpha particles consume only a small fraction
       of the energy and linear momentum fluxes of protons, they cannot be neglected when
       considering the proton angular momentum flux ${\cal L}_p$. 
In most examples, it is found that ${\cal L}_p$ is determined by $v_{\alpha p}$ for
       both the fast and the slow wind. 
In the slow solar wind, it is also found that the proton and alpha angular momentum fluxes ${\cal L}_p$
       and ${\cal L}_\alpha$ can be several times larger in magnitude than the flux carried by the magnetic
       stresses ${\cal L}_M$. 
While the sum of the angular momentum fluxes ${\cal L}_P={\cal L}_p+{\cal L}_\alpha$
       of both species is found to be smaller than the magnetic stress ${\cal L}_M$, for the fast
       and slow wind alike, this result is at variance with the Helios measurements. 
\end{abstract}
\keywords{solar wind --- Sun: magnetic fields}

\section{INTRODUCTION}
The relative velocity between protons and alpha particles
     in the solar wind, ${\bf v}_{\alpha p} \equiv {\bf v}_\alpha -{\bf v}_p$,
     offers important clues for understanding the mechanisms responsible
     for the solar wind acceleration.
In the fast solar wind with proton speeds $v_p =|{\bf v}_p| \gtrsim 600$\velunits,
     Helios measurements
     made in near-ecliptic regions indicate 
     that there exists
     a substantial $v_{\alpha p}=|{\bf v}_{\alpha p}| \mbox{sign}(v_\alpha-v_p)$
     which may be $150$\velunits\ at the heliocentric distance $r \approx 0.3$~AU, 
     amounting to $\sim 1/4$ of the local proton speed.
This $v_{\alpha p}$ decreases with increasing $r$ to
     $30 - 40$\velunits\ at 1~AU.
Furthermore, for a continuous high-latitude fast stream sampled by Ulysses
     between May 1995 and Aug 1996, it was found that the average $v_{\alpha p}$
     decreases from $\sim 40$\velunits\ at 1.5~AU to $\sim 15$\velunits\ at 4.2~AU
     \citep{Reisenfeld_etal_01}. 
As for the slow stream with $v_p\lesssim 400$\velunits,
     $v_{\alpha p}$ tends to be zero  on average
     (see \citet{Marsch_etal_82} and references therein).
However, as clearly shown by Figure~11 in \citet{Marsch_etal_82},  
     there seem to be two categories of slow solar winds, in one
     alpha particles tend to flow faster than protons whereas in the other
     this tendency is reversed.
That $v_{\alpha p} \approx 0$ therefore
     reflects the fact that the two kinds of slow winds have nearly equal opportunities
     to be present.
\citet{Marsch_etal_82} interpreted $v_{\alpha p}>0$
     as a consequence of the wave acceleration that favors alpha particles, 
     and $v_{\alpha p}<0$ as a signature of the solar wind
     being driven primarily by the electrostatic field
     since alpha particles experience only half the electric
     force that protons do.
The former interpretation has been corroborated by an example on day 117 of 1978, 
     during which period there existed simultaneously
     strong Alfv\'en wave activities as well as a significant positive
     $v_{\alpha p}$ at $r\approx 0.29$~AU.
This $v_{\alpha p}$ ($\sim 100$\velunits) is $20-30$\% 
     of the measured proton speed 
     \citep{Marsch_etal_81}.

Helios measurements have also yielded information concerning the angular
      momentum transport in the solar wind, in particular the distribution
      of the angular momentum loss ${\cal L}$ between particles ${\cal L}_P$
      and magnetic stresses ${\cal L}_M$, and the further partition
      of ${\cal L}_P$ between protons ${\cal L}_p$ and alpha particles ${\cal L}_\alpha$
      \citep{Pizzo_etal_83, MarschRichter_84}.
In spite of the significant scatter,
      the data nevertheless exhibit a distinct trend for
      ${\cal L}_p$ to be positive (negative) for solar winds
      with proton speeds $v_p$ below (above) 400\velunits.
The precise measurement of ${\cal L}_\alpha$ is even more difficult, however on average
      ${\cal L}_\alpha$ displays a tendency similar to ${\cal L}_p$.
The magnetic contribution ${\cal L}_M$, on the other hand, is remarkably constant.
A mean value of ${\cal L}_M = 1.6 \times 10^{29}$\specangunits\ can be
      quoted for the solar winds of all flow speeds
      and throughout the region from 0.3 to 1~AU.
For comparison, the mean values of angular momentum fluxes carried by ion flows
      in the slow solar wind are
      ${\cal L}_p = 19.6$ and ${\cal L}_\alpha = 1.3$~$\times 10^{29}$\specangunits\
      (see table II of \citet{Pizzo_etal_83}). 
The fluxes carried by all particles is then 
      ${\cal L}_P = {\cal L}_p + {\cal L}_\alpha= 20.9 \times 10^{29}$\specangunits,
      which tends to be larger than ${\cal L}_M$.
It should be noted that, such a value for ${\cal L}_P$ corresponds to an azimuthal speed
      of more than 7\velunits\ for the bulk slow wind,  consistent  with the
      measurements made before the Helios era
      (see the data compiled in section I of \citet{Pizzo_etal_83}).

The aforementioned problems are not isolated from each other.
As a matter of fact, it has been shown that in the presence of solar rotation,
     an additional force appears in the meridional momentum equation
     of minor ion species.
The effect of this force is to bring the meridional speed of minor ions
     to that of protons
     \citep{McKenzie_etal_79, HI_81}.
It is noteworthy that when using the azimuthal speed of minor ions, 
     \citet{McKenzie_etal_79} were mainly concerned with deriving the so-called
     ``rotational force'' rather than the angular momentum transport.
Hence the azimuthal proton speed was neglected altogether.
\citet{HI_81} offered a more convenient and more self-consistent
     derivation of that force by working in the corotating frame,
     however the azimuthal dynamics was again neglected.
An extension to the study of \citet{McKenzie_etal_79}
     has been recently given by \citet{LiLi_06} (hereafter paper I) who
     worked in the inertial frame and, in improvement of \citet{McKenzie_etal_79},
     treated the protons and alpha particles on an equal footing.
The resultant model is in effect a three-fluid version of the model of
     \citet{WD_67}. 
A low-latitude fast solar wind solution was worked out, and it was shown that the 
     solar rotation introduces a barely perceptible difference in
     the profiles of the meridional ion speeds within 1~AU.
However, the proton-alpha differential streaming plays a decisive role
     in determining the azimuthal speeds, and thus the individual angular momentum fluxes,
     of not only alpha particles but also protons.

{Li \& Li}'s (2006) model  was restricted to the region within 1~AU, and 
     gave only one solution for the purpose of presenting a general analysis
     on the angular momentum transport in a three-fluid solar wind. 
In this study we explore, in a quantitative and systematic manner,
     the interplay between the angular momentum transport and proton-alpha
     differential streaming in both the slow and fast solar wind for the region
     extending from the coronal base out to 4.5~AU.
It should be noted that so far the only available measurements 
     of the alpha angular momentum flux
     are from Helios which explored the near ecliptic solar wind
     between 0.3 and 1~AU \citep{Pizzo_etal_83, MarschRichter_84}.
The Helios measurements will therefore be compared with the computed
     ion angular momentum fluxes for the region within 1~AU only.
The numerical results for the region outside 1~AU
     will concern only the proton-alpha differential streaming $v_{\alpha p}$,
     therefore allowing a comparison of $v_{\alpha p}$
     with the Ulysses measurements.

Before proceeding, we note that
{several mechanisms have been proposed to account for the observed
    evolution of the proton-alpha differential speed $v_{\alpha p}$ in the 
    fast solar wind beyond 0.3~AU.
Since they have been critically reviewed in the introduction section
    of \citet{Kaghashvili_etal_03}, some brief remarks are sufficient here.
First, the pondermotive force due to Alfv\'en waves tends to limit $|v_{\alpha p}|$, 
    but it operates too slowly to reduce $v_{\alpha p}$ in the observed manner,
    especially in the region between 0.3 and 1~AU.
Second, various microinstabilities, in particular the magnetosonic one, are expected
    to be operational when they have a threshold of $v_{\alpha p}$ of the order 
    of the local Alfv\'en speed.
However, these microinstabilities are more likely to be effective
    in regions beyond 1~AU where the parallel proton beta is large.
For a more comprehensive review of these kinetic aspects, 
    please see \citet{Marsch_06}.
Third, the compressional waves accompanying large amplitude Alfv\'en waves may 
   help convert the free energy that derives from the differential streaming
   into alpha heating, thereby reducing $v_{\alpha p}$
   \citep{Kaghashvili_etal_03}.
Finally, although intuitively appealing, the Coulomb friction proves inefficient
     for regulating $v_{\alpha p}$
     in the majority of the solar wind measured by Helios.
It may play some role only for the relatively dense slow wind
     when $|v_{\alpha p}|\lesssim 15$\velunits
     \citep[see Figs.13 and 15 in][]{Marsch_etal_82}.
The functional dependence on $|v_{\alpha p}|$ of the Coulomb collision frequency between
     protons and alpha particles, contained in the coefficient $c_0$ in
     Equation~(\ref{eq_vkl}), dictates that 
     the friction force will rapidly decrease rather than increase
     with increasing $|v_{\alpha p}|$
     if $|v_{\alpha p}|$ exceeds a critical value.
Except for the Coulomb friction, the aforementioned mechanisms are not included
     in the present study as we try
     to isolate the effect of solar rotation
     on the evolution of $v_{\alpha p}$,
     since it is an inherent process whose existence
     is independent of the wave processes.
We further note that}
the effect of solar rotation has
     been incorporated in a number of three-fluid solar wind models
     (e.g., \citet{IH_83}, 
     \citet{BurgiGeiss_86}, \citet{HuHabbal_99}).
However, in all these papers, the spiral magnetic field is assumed rather than
     computed, and the azimuthal speeds of ions are neglected.
These models are therefore not suitable for our purpose, i.e., to examine the 
     angular momentum transport in a three-fluid solar wind.

The paper is organized as follows. 
We first give a brief overview of the model in section~\ref{sec_model}.
Some further details on the numerical implementation of the model are
     given in Section~\ref{sec_nummodel}.
Then section~\ref{sec_numres} presents the numerical results, which are summarized 
     in section~\ref{sec_conc}.

\section{PHYSICAL MODEL}
\label{sec_model}
The solar wind model consists of three species,  electrons ($e$), protons ($p$),
    and alpha particles ($\alpha$).
Each species $s$ ($s=e, p, \alpha$) is characterized by
    its density $n_{s}$,  velocity ${\bf v}_{s}$,
    mass $m_{s}$, electric charge $e_{s}$,
    and temperature $T_{s}$.
The electric charge is also measured in units of electron charge $e$, i.e.,
    $e_{s} = Z_{s} e$ with $Z_e \equiv -1$ by definition.
The ion mass number $A_k$ follows from the relation $m_k = A_k m_p$ ($k=p, \alpha$). 
The mass density of species ${s}$ is $\rho_{s} = n_{s} m_{s}$,
    and the species partial pressure is $p_{s}= n_{s} k_B T_{s}$, where
    $k_B$ is the Boltzmann constant.
From quasi-neutrality it follows that $n_e=n_p+Z_\alpha n_\alpha$.
We also assume quasi zero-current, i.e.,  
    ${\bf v}_{e}= (n_p {\bf v}_p+ Z_\alpha n_\alpha {\bf v}_\alpha)/n_e$,
    except when the ion momentum equations are derived.
The governing equations, which self-consistently take into account the
    effect of solar rotation, are identical to those in paper I
    and will not be given here.
They are appropriate for a time-independent solar wind assuming azimuthal symmetry, i.e., 
    $\partial/\partial \phi \equiv 0$ in a heliocentric spherical
    coordinate system ($r$, $\theta$, $\phi$).
In what follows, we will describe briefly the model, with the emphasis on
    the possible effects introduced by the azimuthal components.
In addition, we will also give an analysis of the distribution
     of the total angular momentum flux between the magnetic field
     and particles.

\subsection{Model Description}
\label{sec_model_dscrpt}
It was shown in the appendix of paper I that no mass or energy exchange between
     different flux tubes is possible, thereby allowing 
     the system of equations to be expressed as a force balance condition
     across the poloidal magnetic field 
     coupled with the conservation equations along it.
In this study, the force balance condition is replaced
     by prescribing a radial (i.e., perfectly monopolar)
     magnetic field.
The model equations can therefore be solved for the radial distribution of
     the densities $n_k$ and radial speeds $v_{k r}$ of
     ion species ($k = p, \alpha$),
     the temperatures $T_{s}$ of all species (${s}=e, p, \alpha$),
     as well as the azimuthal components of the 
     magnetic field $B_\phi$ and ion velocities $v_{k\phi}$.
As such, the model can be seen as a 1.5-dimensional (1.5-D)
     one in that the only independent variable
     is the heliocentric distance $r$, however both the radial and azimuthal components of
     vectors are retained.
To illustrate the effects of solar rotation, we will also compute
     models without the $\phi$ components.
For the ease of description these models will be called 1-D ones.
Furthermore, the ion heat fluxes are neglected, 
     and the field-aligned  electron heat flux ${\bf q}_e$
     is assumed to follow the Spitzer law, 
     which results in $\nabla\cdot{\bf q}_e = -(1/a)(\partial/\partial r)
     [a\kappa T_e^{5/2}(\partial T_e/\partial r) \cos^2\Phi]$.
Here the cross-sectional area of the flux tube $a$ scales as
      $a \propto 1/B_r \propto r^2$,
      and the magnetic azimuthal angle $\Phi$
      is defined by $\tan\Phi=B_\phi/B_r$.
As for the electron conductivity $\kappa$, the Spitzer value is used,
      $\kappa = 7.8\times 10^{-7}$ {erg}~{K}$^{-7/2}$~{cm}$^{-1}$~s$^{-1}$
      \citep{Spitzer_62}.

From paper I, the equation governing the radial speed of ion species $k$ is
\begin{eqnarray}
 v_{k r}\frac{\partial v_{k r}}{\partial r}
    &=& -\frac{1}{n_k m_k} \frac{\partial p_k}{\partial r} 
       -\frac{Z_k}{n_e m_k}\frac{\partial p_e}{\partial r}  
       - \frac{G M_\odot}{r^2}  \nonumber \\
    &+& a_{k} + \frac{n_j}{A_k n_e}c_0(v_{j r}-v_{k r})\sec^2\Phi \nonumber \\
    &+& \left[\frac{v_{k\phi}^2}{r}
        -\tan\Phi v_{k r}\left(\frac{\partial v_{k\phi}}{\partial r} 
        +\frac{v_{k\phi}}{r} \right) \right]  , 
        \label{eq_vkl} 
\end{eqnarray}
     where the subscript $j$ denotes the species other than $k$, i.e., 
     $j=\alpha$ for $k=p$ and vice versa,
     and $c_0$
     is a coefficient associated with Coulomb frictions.
Moreover, $G$ is the gravitational constant, $M_\odot$ is the mass of the Sun,
     and $a_k$ stands for the acceleration from some external process. 
Apart from $a_{k}$,  the meridional acceleration of the ion flow
     includes the ion and electron pressure gradient forces
     (the terms proportional to $\partial p_k/\partial r$ and 
      $\partial p_e/\partial r$, respectively),
     the gravitational force,
     the Coulomb friction corrected for the spiral magnetic field,
     and the force associated with azimuthal flow speeds (the term in the square parentheses,
     hereafter referred to as the ``azimuthal force'' for brevity).
Note that the electron pressure gradient force is part of the electrostatic
     force which is apportioned between two ion species according to their charge-to-mass ratios
     $Z_k/A_k$ ($k=p, \alpha$).

Introducing $\phi$ components impacts the meridional dynamics through two means.
First, the electron conductivity is in effect reduced by a factor $\cos^2\Phi$, thereby
     affecting the electron temperature and hence the electron pressure
     gradient force.
Second, the azimuthal force enters into the meridional momentum equations.
The two effects can both be important, as will be illustrated by the numerical solutions 
     presented in Section \ref{sec_num_slow}.

\subsection{Angular Momentum Loss Rate and Its Distribution between Particles
      and Magnetic Stresses}
\label{sec_model_anganalysis}

In what follows, a simple analysis based on constants of motion
      for the governing equations shows 
      how the angular momentum loss rate per steradian
      ${\cal L}$ is distributed between particles ${\cal L}_P$  and
      magnetic stresses ${\cal L}_M$
      in a solar wind where 
      different ion species flow with different velocities.
As such, this discussion complements that of \citet{WD_67} 
      and \citet{MarschRichter_84} where the solar wind is seen as a bulk flow.

Following paper I, the expressions for the azimuthal components $v_{p \phi}$, $v_{\alpha \phi}$
    and $B_\phi$ are
\begin{subequations}
\label{eq_bvphi} 
\begin{eqnarray}
v_{p\phi} &=& \frac{\Omega r\sin\theta}{M_T^2-1}
   \left[ 
     \overbrace{M_T^2\frac{r_A^2}{r^2}-1}^{\mbox{I}}  
    +\overbrace{M_\alpha^2\frac{v_{\alpha p, r}}{v_{\alpha r}}
         \left(1-\frac{r_A^2}{r^2}\right)}^{\mbox{II}}
   \right],
         \label{eq_vpphi}\\
v_{\alpha\phi} &=& \frac{\Omega r\sin\theta}{M_T^2-1}
   \left[ 
     \overbrace{M_T^2\frac{r_A^2}{r^2}-1}^{\mbox{I}} 
    +\overbrace{M_p^2\frac{v_{\alpha p, r}}{v_{p r}}
         \left(\frac{r_A^2}{r^2}-1\right)}^{\mbox{II}}
   \right],
         \label{eq_viphi}\\
B_{\phi} &=& \Omega r\sin\theta\frac{4\pi\rho_p v_{p r} (1+\zeta)}{B_r}
   \frac{r_A^2/r^2-1}{M_T^2-1}, 
         \label{eq_bphi}
\end{eqnarray}
\end{subequations}
    in which $v_{\alpha p, r}$ is the radial component of
    the velocity difference vector ${\bf v}_{\alpha p}$,
    $\Omega$ is the angular rotation rate of the flux tube,
    and the constant $\zeta=(\rho_\alpha v_{\alpha r})/(\rho_p v_{p r})$
    denotes the ion mass flux ratio.
The terms designated by II are associated with the differential streaming,
    whereas the terms denoted by I take care of the rest.
The subscript $A$ denotes the Alfv\'enic point where
    $M_T=1$, $M_T$ being the combined meridional
    Alfv\'enic Mach number defined by
\begin{eqnarray}
M_T^2  = M_p^2+M_\alpha^2, \hskip 0.2cm 
M_k^2=\frac{v_{k r}^2}{B_r^2/4\pi\rho_k}, 
\label{eq_def_mach}
\end{eqnarray}
    with $k=p,\alpha$.

The field and particle contributions to the angular momentum
    loss rate per steradian ${\cal L}$ are 
\begin{eqnarray}
{\cal L}_M = - r^3\sin\theta\frac{B_\phi B_r}{4\pi}, \hspace{0.2cm}
{\cal L}_k = r^3\sin\theta \rho_k v_{k r} v_{k\phi}, 
\end{eqnarray}
    where $k=p, \alpha$.
From Equation~(\ref{eq_bvphi}) follows that
\begin{eqnarray}
{\cal L} = \dot{M}\Omega r_A^2 \sin\theta_A^2 ,
\label{eq_calL}
\end{eqnarray}
    in which $\dot{M}=(1+\zeta) \rho_p v_{p r} r^2$ is the total mass loss rate
    of the solar wind.
The ratio ${\cal L}_k/{\cal L}_M$ is rather complex and had better be examined 
    case by case
{due to the presence of terms labeled II in Equations~(\ref{eq_vpphi})
    and (\ref{eq_viphi}).
Note that terms II derive essentially from the requirement that the 
    proton-alpha velocity difference vector be aligned with the instantaneous
    magnetic field \citep{LiLi_06}.
The azimuthal speeds are expected to be determined by these terms when there exists
    a substantial $v_{\alpha p, r}$, which is almost exclusively determined
    by the way in which the external energy is distributed among different species
    in the corona.
As shown in section~\ref{sec_numres}, such a situation happens for the majority
    of the numerical solutions in interplanetary space.
Hence the azimuthal speeds and the resultant specific angular momenta of ion species
    measured {\it in situ} are essentially determined by the processes that happen
    in the corona.
Furthermore, it follows from the identity
    $\rho_p v_{pr} M_\alpha^2/v_{\alpha r} = \rho_\alpha v_{\alpha r} M_p^2/v_{p r}$
    that terms II have no contribution to the overall angular momentum loss rate.
As a result,}
    a concise result can be found  for the ratio of the overall particle 
    contribution ${\cal L}_P={\cal L}_p+{\cal L}_\alpha$
    to the magnetic one ${\cal L}_M$.
By defining the bulk velocity 
\begin{eqnarray}
{\bf u}= ({\bf v}_p + \zeta {\bf v}_\alpha)/(1+\zeta),
\label{eq_def_bulku}
\end{eqnarray}
     one can see $M_T^2 = (r/r_A)^2 (u_r/u_{r, A})$.
As a result, 
\begin{eqnarray}
\frac{{\cal L}_P}{{\cal L}_M}
   =  \frac{u_r/u_{r,A}-1}
      {1-r_A^2/r^2}.
\label{eq_distriL2}
\end{eqnarray}
It is interesting to see that only one single speed $u_r$ enters into
     the ratio ${\cal L}_P/{\cal L}_M$.

In comparison with \citet{WD_67} and \citet{MarschRichter_84}, the expressions here
    indicate that, a proper definition of the Alfv\'enic point
    is through $M_T=1$, and the bulk flow velocity ${\bf u}$ should be defined
    as an average between the proton and alpha velocities, weighted by their
    respective mass fluxes (cf. Eq.(\ref{eq_def_bulku})).
Of course these differences only matter when the abundance of alpha particles 
    and the differential streaming between the protons and alpha particles 
    are large. 
When $\zeta =0$ or ${\bf v}_\alpha = {\bf v}_p$,
    the results are identical to those in \citet{MarschRichter_84}.

\section{NUMERICAL IMPLEMENTATION}
\label{sec_nummodel}

The model has been briefly described in the preceding section where a rather general analysis
     is also given.
However, it is still necessary to solve the governing equations numerically for 
     a quantitative analysis to be made.
In particular, some external heating and/or momentum addition need to be applied to
     the ions to generate a solar wind solution.
The implementation of the energy deposition and a 
     description of the method of solution are given in this section.

\subsection{Energy Deposition}
\label{sec_nummodel_assumphting}
The external energy deposition is assumed to come from
      an {\it ad hoc} energy flux which
      dissipates at a constant length $l_d$. 
The resultant total heating rate is therefore 
\begin{eqnarray*}
Q=F_{E} \frac{B_r}{B_{r E}l_d} \exp\left[-\frac{(r-R_\odot)}{l_d}\right],
\end{eqnarray*}
      where $F_E$ is the input flux scaled to
      the Earth orbit $R_E = 1$~AU, 
      $B_{r E}=3.3\gamma$ is the radial magnetic field strength at $R_E$,
      and $R_\odot$ is the solar radius.
The energy is deposited in protons entirely as heat $Q_p$,
      but deposited in the alpha gas in the form of both heating $Q_\alpha$
      and acceleration $a_\alpha$.
To be more specific, $Q$ is apportioned among $Q_p$, $Q_\alpha$
      and $a_\alpha$ in the following way,
\begin{subequations}
\begin{eqnarray}
&& Q_p + \bar{Q}_\alpha = Q, \hspace{0.2cm}
\frac{\bar{Q}_\alpha}{Q_p} = \frac{\chi \rho_\alpha}{\rho_p},   \\
&& Q_\alpha + \rho_\alpha v_{\alpha r} a_\alpha =\bar{Q}_\alpha ,\hspace{0.2cm}
  \frac{a_\alpha}{Q_\alpha} = \frac{\chi_d}{\rho_\alpha v_0}, \label{eq_chi}
\end{eqnarray}
\end{subequations}
     where $v_0 = \sqrt{k_B T_p/m_p}$ is some characteristic speed.

{The choice of the heat and momentum deposition needs some explanation.
The exponential form of $Q$ was first suggested by \citet{HolzerAxford_70}
    and was later employed in a large number of studies.
The form of $Q$ adopted here is slightly different from the original version to ensure that
    it mimics the dissipation of a flux of non-thermal energy, for instance,
    the dissipation of low-frequency Alfv\'en waves.
This may happen as a result of 
    a turbulent cascade towards high parallel wavenumbers where
    the wave energy is picked up by ions through the ion-cyclotron resonance.
(For more details, please see the review by \citet{HI_02} dedicated
    to this topic.)
Once $Q$ is specified, the way it is distributed among different
    ion species, characterized by the parameter $\chi$, depends on
    the resonant interaction between cyclotron waves and ions.
Note that $\chi$ indicates how alpha particles
      are favored when $Q$ is distributed, with
      $\chi=1$ standing for the neutral heating:
      the total energy that goes to ion species $k$ is proportional
      to its mass density $\rho_k$ ($k=p, \alpha$).
Previous computations involving ion cyclotron waves indicate that, 
      in the case of neutral heating, the alpha particles
      tend to flow slower than protons 
      \citep[see the dispersionless case in Fig.1 of][]{HuHabbal_99}.
Only when the alphas are energetically favored in the corona
      can the modeled
      $v_{\alpha p}$ be positive in interplanetary space.
This happens when $\chi >1$.
(Note that we adopted a constant $\chi$ throughout
      the computational domain for simplicity.)
 }
Listed in Table~\ref{tbl_1}, and unless otherwise stated, 
      {the heating} parameters
      are chosen to produce fast or slow solar wind solutions with realistic
      ion mass fluxes and terminal speeds.

\begin{deluxetable}{c c c c c}
\tablewidth{0pt}
\tablecaption{
Parameters characterizing the energy deposition to ion species
\label{tbl_1}
}
\tablehead{
wind	& $F_E$				& $l_d$		&$\chi$		& $\chi_d$ \\
type	& (erg cm$^{-2}$ s$^{-1}$)	& ($R_\odot$)	& & }
\startdata 
Fast & 1.8 & 5 & 1.8 & see note\tablenotemark{a} \\
Slow & 1.2 & 1 & 2.9 & 0 \\ 
\enddata
\tablenotetext{a}{$\chi_d$ decreases with $r$ from $30$ 
    to $1$ with a sharp transition at 2~$R_\odot$}
\end{deluxetable}

\subsection{Method of Solution and Boundary Conditions}
\label{sec_nummodel_solmethod}
The governing equations (Eqs.(1) to (7) in paper I) are cast
    in a time-dependent form
    and are solved by using a fully implicit numerical scheme
    \citep{Hu_etal_97}.
Starting from an arbitrary guess, the equations
    are advanced in time until a steady state is reached.
The computational domain extends from the coronal base (1~$R_\odot$) to
    4.5~AU.
At the base, the ion densities and species temperatures are fixed,
    $n_e=3\times 10^8\mbox{~cm}^{-3}$, $n_\alpha/n_p = 0.1$, 
    $T_e=T_p=T_\alpha=1.2 \times 10^6\mbox{~K}$.
The radial components of ion velocities, $v_{p r}$ and $v_{\alpha r}$, are specified
    to ensure mass conservation.
On the other hand, $v_{p\phi}$ and $B_\phi$ are specified according to
    Equation~(\ref{eq_bvphi}).
At the outer boundary ($4.5$~AU), all dependent variables
    are linearly extrapolated for simplicity.
We take $\Omega=2.865\times 10^{-6}$~rad s$^{-1}$, which corresponds to
    a sidereal rotation period of $25.38$ days.

\section{NUMERICAL RESULTS}
\label{sec_numres}

As has been described in the introduction, we are interested in
     answering two questions:
     1) To what extent is $v_{\alpha p}$, 
            the differential streaming between protons and alpha particles,
            affected by introducing
            solar rotation? and
     2) Is it possible to reconcile the model results 
           with the Helios measurements, as far as the specific 
           angular momentum fluxes are concerned? 
In this section, we will first examine the fast solar wind and then
     move on to the slow one.

\begin{figure*}
\epsscale{.9}
\plotone{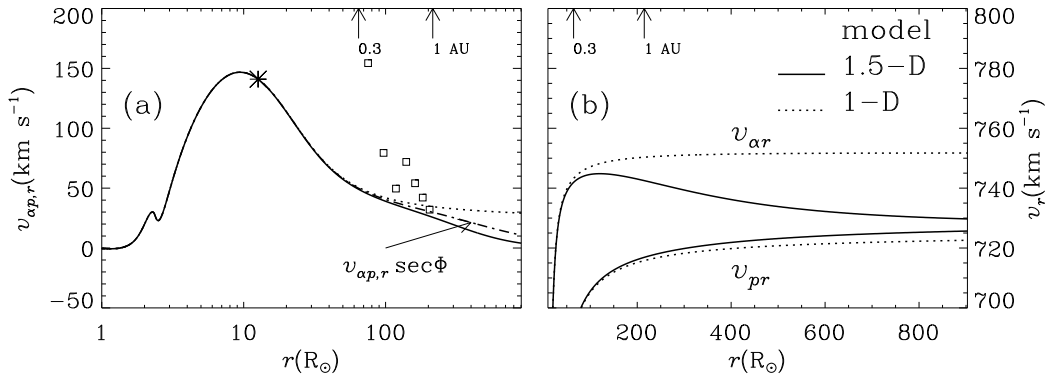}
\caption{
    Effect of solar rotation on a 3-fluid fast solar wind in the equatorial
          plane $\theta=90^\circ$.
    The azimuthal components of ion velocities and the magnetic field
          are incorporated self-consistently in the 1.5-D model (solid curves),
          but excluded in the 1-D model (dotted curves).
    Radial profiles are shown for several flow parameters.
    (a) Radial component of the proton-alpha relative velocity
            ${\bf v}_{\alpha p} = {\bf v}_{\alpha} - {\bf v}_{p}$,
    and (b) the radial speeds of protons $v_{p r}$ and alpha particles $v_{\alpha r}$.
    In panel (a), ${v}_{\alpha p} = v_{\alpha p, r} \sec\Phi$ is given by
          the dash-dotted line, where $\Phi$
         is the magnetic azimuthal angle, defined by $\tan\Phi=B_\phi/B_r$.
    The asterisk refers to the Alfv\'enic point where $M_T=1$, $M_T$ being the
         combined Alfv\'enic Mach number defined by Equation~(\ref{eq_def_mach}).
    In addition, the open boxes give the near-ecliptic Helios measurements of $v_{\alpha p}$
         reported by \citet{Marsch_etal_82}.
    }
\label{fig_cpfast}
\end{figure*}

\subsection{Fast Solar Wind Solutions}
\label{sec_num_fast}

\subsubsection{Effect of Solar Rotation on the Differential Streaming}
\label{sec_num_fast_vap}

Given in Figure~\ref{fig_cpfast} is a comparison of the 1.5-D (solid curves)
     with the 1-D model (dotted curves) 
     of the fast solar wind solution in the equatorial plane (colatitude $\theta=90^\circ$).
Plotted are the radial profiles for
     (a) the radial component of the proton-alpha velocity difference
                $v_{\alpha p, r}$,
     and 
     (b) the radial ion speeds $v_{p r}$ and $v_{\alpha r}$.
In Fig.\ref{fig_cpfast}a, the dash-dotted line shows
     the parameter $v_{\alpha p} = v_{\alpha p, r}\sec\Phi$, which
     also takes into account the azimuthal component of the difference vector. 
The asterisk denotes the Alfv\'enic point in the 1.5-D model, which lies 
     at $r_A=12.6 R_\odot$.
For comparison, the open boxes give the near-ecliptic Helios measurements of $v_{\alpha p}$
     for the fast wind as reported by \citet{Marsch_etal_82}.

From Figs.\ref{fig_cpfast}a and \ref{fig_cpfast}b, one can see that
      introducing the solar rotation only affects the solar wind beyond say 50~$R_\odot$.
Thus it is not surprising that the ion fluxes are unchanged
      in the 1.5-D model, as compared with the 1-D one.
Both models yield an ion flux ratio of
      $(n_\alpha v_{\alpha r})/(n_p v_{p r}) = 0.026$,
      and a proton flux of $(n_p v_{p r})_E = 2.72\times 10^8$~cm$^{-2}$ s$^{-1}$
      when scaled to 1~AU.

It can be seen from Fig.\ref{fig_cpfast}a that, 
      although $v_{\alpha p, r}$ decreases with radial distance $r$
      for both the 1-D and 1.5-D models beyond their mutual maximum
      of $147$\velunits\ attained at 9.4~$R_\odot$, 
      the reduction in $v_{\alpha p, r}$ is more prominent
      in the 1.5-D model from 0.3 to 4~AU.
The reduction in $v_{\alpha p}$ for the 1.5-D model, when compared with
     the 1-D one, is also substantial, although less significant
     than the reduction in $v_{\alpha p, r}$ due to the presence of
     the azimuthal component of ${\bf v}_{\alpha p}$.
Nevertheless the 1.5-D model yields a $ v_{\alpha p}$ of 11.4\velunits\ at 4~AU,
     as opposed to 29.3\velunits\ attained in the 1-D model.
From Fig.\ref{fig_cpfast}b one can see that the reduction in $v_{\alpha p, r}$
     in the 1.5-D model is achieved by lowering the radial speed
     profile of alpha particles $v_{\alpha r}$
     and raising that of protons $v_{p r}$.
This effect is barely perceptible at 0.3~AU but becomes more obvious with increasing $r$.

\begin{figure*}
\epsscale{.9}
\plotone{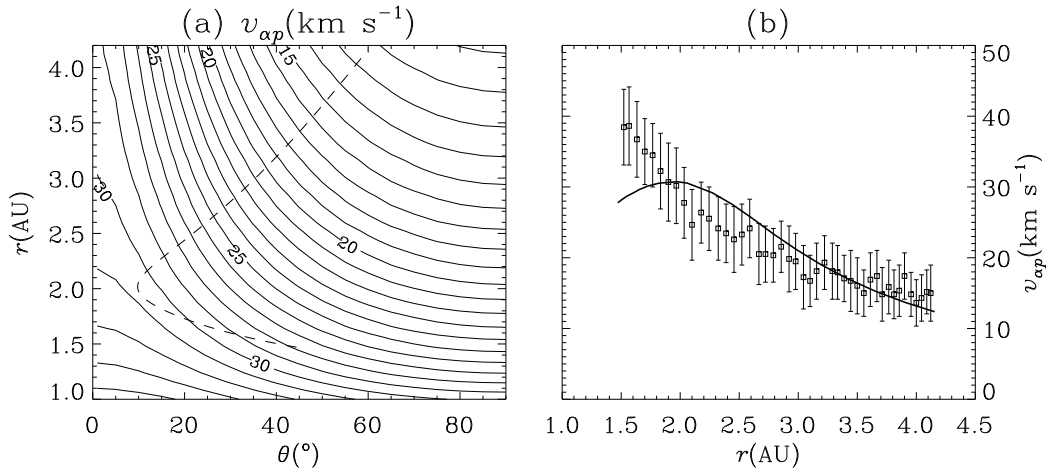}
\caption{Effect of solar rotation on a 3-fluid fast solar wind.
(a) Distribution in the $r-\theta$ plane of the proton-alpha differential
      speed $v_{\alpha p}$ for the region extending from $1$ to $4.2$ AU.
The contours are equally spaced by 1\velunits.
The dashed line depicts the trajectory of Ulysses in the interval between
    1995 May 05 and 1996 Aug 02.
The solutions are obtained by varying the colatitude $\theta$ alone with all other parameters unchanged.
As such, the latitudinal dependence of $v_{\alpha p}$ is entirely due to solar rotation.
(b) Radial profile of the modeled $v_{\alpha p}$ along the Ulysses trajectory (solid line). 
The error bars represent the 25-75\% percentiles of the $v_{\alpha p}$ distribution
     over 10-day bins as sampled by Ulysses \citep{Reisenfeld_etal_01}.
    }
\label{fig_cpUly}
\end{figure*}

Figure~\ref{fig_cpfast}a indicates that the modeled $v_{\alpha p}$ profile 
     deviates significantly from the Helios measurements,
     especially in the region closer to the Sun. 
In other words, the deceleration of alpha particles relative to protons
     as measured by Helios cannot be explained by merely invoking
     the $\phi$ components.
However can this mechanism be entirely ruled out as far as the Ulysses
     measurements are concerned?
Such a possibility is examined in Figure~\ref{fig_cpUly}.
In Fig.\ref{fig_cpUly}a, the distribution of $v_{\alpha p}$ in the $r-\theta$ plane for the region
     extending from 1 to 4.2~AU is displayed as contours equally spaced by 1\velunits.
The solutions are obtained by varying the colatitude $\theta$ alone.
Recalling that by assumption the solar wind
     flows in a perfectly monopolar magnetic field,
     one may see that the latitudinal dependence is entirely due to solar rotation.
From Fig.\ref{fig_cpUly}a it is obvious that in the regions close to the pole,
     say $\theta \lesssim 10^\circ$, 
     $v_{\alpha p}$ shows little radial dependence.
However with increasing $\theta$, the radial gradient
     in $v_{\alpha p}$ becomes increasingly significant.

In Fig.\ref{fig_cpUly}b, the solid curve represents the modeled $v_{\alpha p}$ profile
     along the Ulysses trajectory delineated by the dashed curve in Fig.\ref{fig_cpUly}a
     in the interval between 1995 May 05 and 1996 Aug 02.
During this period, Ulysses sampled a continuous, undisturbed, high speed stream
     above the latitude of $30^\circ$N in the radial range of $1.5\sim 4.2$~AU.
The actual Ulysses measurements are given by the error bars
     which correspond to the 25-75\% percentiles of the $v_{\alpha p}$ distribution
     over 10-day bins 
     \citep{Reisenfeld_etal_01}.
It can be seen that, for $r \lesssim 2$~AU, $v_{\alpha p}$ in the numerical solutions increases
     as opposed to the observed tendency for $v_{\alpha p}$ to decrease.
This is because during this period, Ulysses was moving towards the pole and the effect of
     solar rotation diminishes with decreasing $\theta$.
However, for $r \gtrsim 2$~AU, the decrease in $v_{\alpha p}$ from 30.7\velunits\
     at 2~AU to 12.4\velunits\ at 4.2~AU is consistent with the observed values
     within the accuracy of the measurements.
From this we conclude that solar rotation should not be neglected in
     attempts to understand the Ulysses measurements
     of $v_{\alpha p}$ beyond 2~AU.
It is necessary to stress that this conclusion should not be confused with that reached by    
     \citet{Reisenfeld_etal_01}
     who demonstrated by showing a profile designated ``Ulysses rotational deceleration''
     in their Figure~8 that the $v_{\alpha p}$ profile measured by Ulysses has
     little to do  with the rotational deceleration.
In fact, we agree with these authors that introducing solar rotation cannot 
     provide a unified mechanism  to account for the observed
     deceleration of $v_{\alpha p}$ from 0.3~AU onwards.
What Fig.\ref{fig_cpUly}b shows is that, if a reasonable $v_{\alpha p}$ is given
     at 2~AU, the solar rotation alone can decelerate the alpha particles relative to protons
     in the observed fashion.

\subsubsection{Angular Momentum Transport}
\label{sec_num_fast_angmom}
\begin{figure}
\begin{center}
\epsscale{.8}
\plotone{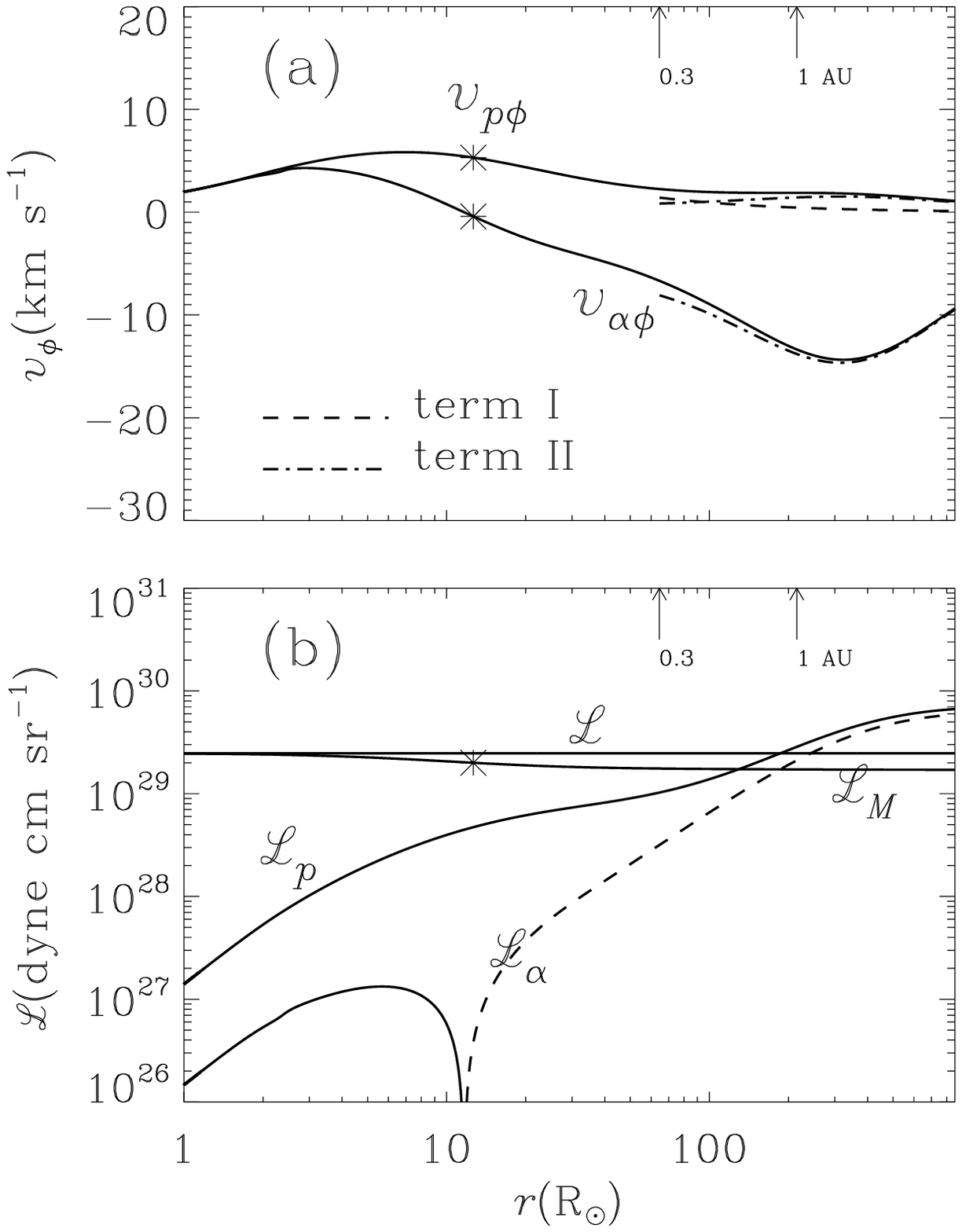}
\caption{Angular momentum transport in a 3-fluid fast solar wind.
    (a) Azimuthal speeds of protons $v_{p\phi}$ and alpha particles $v_{\alpha \phi}$,
    and (b) specific angular momentum fluxes carried by protons ${\cal L}_p$,
            alpha particles ${\cal L}_{\alpha}$
            and magnetic stresses ${\cal L}_M$.
    The fluxes add up to ${\cal L}$.
    In panel (a), the dashed line shows the contribution from term I
        in Eqs.(\ref{eq_vpphi}) and (\ref{eq_viphi}),
        whereas the dash-dotted curves give the contribution from term II.
    In panel (b), the dashed line represents negative values.
    The asterisks mark the position of the Alfv\'enic point $r_A$.
}
\label{fig_azi_spec}
\end{center}
\end{figure}

We now turn to the problem of the angular momentum transport
     in the fast solar wind along $\theta=90^\circ$.
Figure~\ref{fig_azi_spec} describes the radial distribution of
     (a) the azimuthal ion speeds $v_{p\phi}$ and $v_{\alpha\phi}$,
     and (b) the specific angular momentum fluxes including
     those carried by protons ${\cal L}_p$, alpha particles
     $\mathcal{L}_\alpha$ and magnetic stresses ${\cal L}_M$.
The total angular momentum flux ${\cal L}$ should be a constant, as 
     is indeed the case in the numerical solution.
To isolate the contribution of the differential streaming to individual 
     azimuthal speeds, term I (II) in Eqs.(\ref{eq_vpphi}) and (\ref{eq_viphi})
     is plotted as the dashed (dash-dotted) curve
     between 0.3 and 4~AU.
In Fig.\ref{fig_azi_spec}b, the dashed line is used for plotting negative values.
The asterisks in both panels mark the Alfv\'enic point $r_A$.

From Fig.\ref{fig_azi_spec}a one can see that with the 
     development of $ v_{\alpha p,r}$, some substantial difference
     arises between $v_{p\phi}$ and $v_{\alpha\phi}$.
To be more specific, $v_{p \phi}$ increases from the coronal base
     to 5.84\velunits\ at 6.9~$R_\odot$, beyond which $v_{p \phi}$
     decreases gradually to $1.1$\velunits\ around 4~AU.
On the other hand, $v_{\alpha \phi}$ reaches a local maximum of 4.3\velunits\
     at 2.88~$R_\odot$.
Rather than partially corotate with the Sun, alpha particles then gradually
     develop an azimuthal velocity in the direction of counter-rotation
     with the Sun: $v_{\alpha\phi}$ is negative beyond 11.7~$R_\odot$.
Then $v_{\alpha\phi}$ attains a local minimum of -14.4\velunits\ at 324~$R_\odot$
     and increases thereafter to -9.35\velunits\ at 4~AU.
It can be seen that, for $r \ge 0.3$~AU not only $v_{\alpha \phi}$ but also
     $v_{p\phi}$ are determined mainly by term II (the dash-dotted lines)
     in their respective expressions (\ref{eq_vpphi}) and (\ref{eq_viphi}).

Despite the small values of $v_{p\phi}$, 
     Fig.~\ref{fig_azi_spec}b reveals that ${\cal L}_p$
     exceeds ${\cal L}_M$ in the region beyond 129~$R_\odot$.
Furthermore, ${\cal L}_\alpha$ becomes larger than ${\cal L}_M$ in magnitude
     for $r\ge 186$~$R_\odot$.
Consider the values at 1~AU.
As part of the total flux ${\cal L}=2.47$ (all values are in units of $10^{29}$\specangunits),
     the magnetic part ${\cal L}_M = 1.72$ is smaller than the proton contribution 
     ${\cal L}_p = 2.86$ which however is largely
     offset by the alpha contribution ${\cal L}_\alpha = -2.12$.
As a matter of fact, magnetic stresses are always the primary constituent
     in ${\cal L}$, the particle contribution as a whole is 
     smaller than ${\cal L}_M$ throughout the computational domain.
At 1~$R_\odot$, 99.37\% of ${\cal L}$ is contained in magnetic stresses ${\cal L}_M$.
Although ${\cal L}_M$ decreases monotonically with increasing distance, 
     it still amounts to 69.2\% of ${\cal L}$ at 4~AU.
This is understandable in view of Equation~(\ref{eq_distriL2}).
For $r\sim R_\odot$, one can see that $r \ll r_A$ and $u_r \ll u_{r,A}$.
It then follows that ${\cal L}_M/{\cal L}\approx 1-(r/r_A)^2 \approx 1$.
On the other hand, for $r\gg r_A$ one can see ${\cal L}_M/{\cal L}\approx u_{r,A}/u_r$.
That ${\cal L}_M$ takes up most of the total flux ${\cal L}$ asymptotically
     therefore
     stems from the fact that $u_{r,A}/u_r$ varies little beyond $r_A$.

The value of ${\cal L}_M$ (in units of  $10^{29}$\specangunits)
     shows only a slight decrease from 1.76 at 0.3~AU
     to 1.72 at 1~AU.
This is consistent with the quoted value of $1.6 \sim 1.9$ as measured by Helios
     \citep{Pizzo_etal_83, MarschRichter_84}.
However, Helios measurements indicate that in the fast solar wind, 
     both ${\cal L}_\alpha$ and ${\cal L}_p$ tend to be negative.
As a result, the overall particle contribution ${\cal L}_P$ is negative, 
     which can be realized only if the solar wind bulk speed $u_r$ decreases with $r$
     in the region $r\ge r_A$.
This is not possible by simply varying the heating parameters in the model.
Instead, we would rather interpret Fig.\ref{fig_azi_spec} as an indication
     that the the proton angular momentum flux cannot be studied on its own,
     i.e., by simply incorporating $\phi$ components into models of an
     electron-proton plasma.
Although alpha particles take up only a minor part of the solar wind
     momentum and energy fluxes, the proton-alpha differential speed is the primary
     contributor to individual azimuthal ion speeds and therefore
     individual ion angular momentum fluxes.

\subsection{Slow Solar Wind Solutions}
\label{sec_num_slow}
The effect of solar rotation on individual ion speeds in the fast solar wind 
     is rather weak inside 1~AU.
Is this also the case for the slow wind?
Moreover, the Helios measurements show that in the slow solar wind,
     the proton and alpha angular momentum fluxes are both positive.
When added together, they are also larger than that carried by 
     magnetic stresses.
Can this measurement be explained by the present model?
To answer these questions, we will first examine an example
     with a significant positive
     $v_{\alpha p, r}$ to gain some insight.
A parameter study surveying the parameter $\chi$ will then be presented.

\subsubsection{Effect of Solar Rotation on the Differential Streaming}
\label{sec_num_vap_slow}

\begin{figure*}
\epsscale{.9}
\plotone{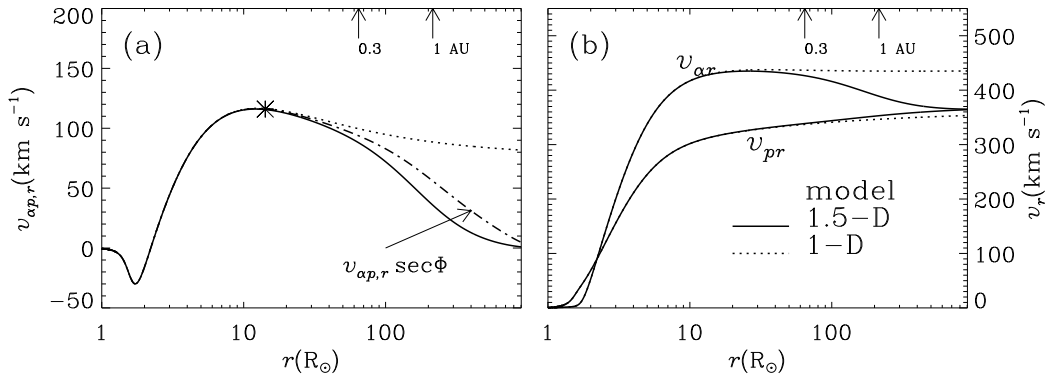}
\caption{
    Effect of solar rotation on 3-fluid slow solar winds in the equatorial
          plane $\theta=90^\circ$.
    The 1.5-D model is given by solid curves whereas the 1-D one
          is given by dotted curves.
    (a) Radial component of the proton-alpha relative velocity
            ${\bf v}_{\alpha p}$
    and (b) The radial speeds of protons $v_{p r}$ and alpha particles $v_{\alpha r}$.
    In panel (a), the asterisk refers to the Alfv\'enic point, and
         the dash-dotted line shows ${v}_{\alpha p}  = v_{\alpha p, r} \sec\Phi$.
    }
\label{fig_cpslow}
\end{figure*}

Figure~\ref{fig_cpslow} compares a 1.5-D (solid curves)
     with 1-D model (dotted curves) 
     of the slow solar wind in the equatorial plane $\theta=90^\circ$.
Plotted are the radial profiles for
     (a) the radial component of the proton-alpha velocity difference
         $v_{\alpha p, r}$, 
     and (b) the radial speeds of ions $v_{k r}$ ($k=p, \alpha$).
In Fig.\ref{fig_cpslow}a, 
     the asterisk denotes the Alfv\'enic point in the 1.5-D model, which 
     now lies at $r_A=14.2$~$R_\odot$.
The dash-dotted line is used to show the parameter $v_{\alpha p}$. 

As in the case for the fast wind, introducing the solar rotation
      does not alter the ion fluxes.
When scaled to 1AU, both the 1-D and 1.5-D models yield
      a proton flux of 
      $(n_p v_{p r})_E = 3.36\times 10^8$~cm$^{-2}$ s$^{-1}$,
      and a flux ratio of
      $(n_\alpha v_{\alpha r})/(n_p v_{p r}) = 0.024$.
Furthermore, introducing the $\phi$ components reduces the proton-alpha differential speed
      by decelerating the alphas and accelerating the protons.
Take the radial speeds at 1~AU for instance.
The 1.5-D model yields $v_{\alpha r} = 389$\velunits, 
     whereas the 1-D model results in $v_{\alpha r} = 435$\velunits.
As for $v_{p r}$, the 1.5-D (1-D) model yields $353$ ($347$)\velunits.
As a result, $v_{\alpha p, r}$ at 1~AU 
     is $88$\velunits\ for the 1-D but $36$\velunits\ for the 1.5-D model.
Going further to 4~AU, in the 1-D model alpha particles still flow considerably 
     faster than protons: values of
     $v_{\alpha r}=435$ and $v_{p r}=353$\velunits\ are found at 4~AU. 
In contrast, $v_{\alpha r}$ and $v_{p r}$ attain nearly the same 
     value ($\sim 365$\velunits) in the 1.5-D model.
It should be noted that, a value of $v_{\alpha p}=88$\velunits\ at 0.3~AU
     seen in Fig.\ref{fig_cpslow}a for the 1.5-D model is not unrealistic.
Even larger values have been found for $v_{\alpha p}$ 
     by Helios 2 when approaching perihelion
     \citep{Marsch_etal_81}.
However $v_{\alpha p}$ at 1~AU seems somehow unrealistic:
     A value of $56$\velunits\ is substantially larger than 
     the local Alfv\'en speed
     $V_A = B_r \sec\Phi/\sqrt{4\pi(\rho_p+\rho_\alpha)}$ which yields 35\velunits.
By contrast, the Helios measurements indicate
     that in the slow wind the ratio $v_{\alpha p}/V_A$ spans
     a broad distribution between
     $-1$ and $1$ but peaks at $\sim \pm 0.25$
     (Figure~11 in \citet{Marsch_etal_82}).
The discrepancy indicates once again that 
     invoking the solar rotation alone is not an efficient way
     to decelerate the alpha particles relative to protons.
However, if some yet unknown mechanism results in a substantial $v_{\alpha p}$ at 0.3~AU
     as observed,
     the effect of solar rotation on the evolution of $v_{\alpha p}$ with radial distance $r$
     beyond 0.3~AU probably cannot be neglected from the perspective of slow solar wind modeling.
For instance, in this particular solution, the changes introduced in the alpha speed
     $v_{\alpha r}$ at the distance of 4 (1)~AU can be 16.1\% (10.5\%) 
     the values in the case without taking solar rotation into account.

\begin{figure}
\begin{center}
\epsscale{.99}
\plotone{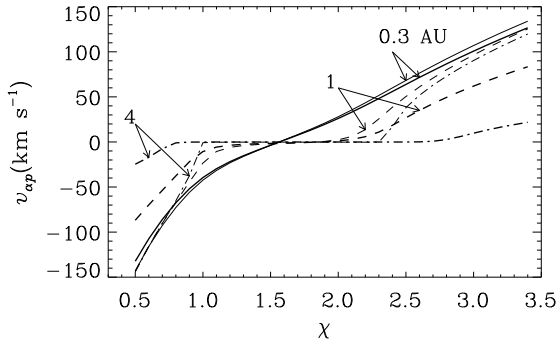}
    \caption{
    Dependence of the proton-alpha differential speed $v_{\alpha p}$ on the parameter $\chi$ 
           at three radial distances, as labeled, 
           for the 1.5-D model (thick curves) and the 1-D one (thin curves).
     }
\label{fig_vap_par}
\end{center}
\end{figure}

Returning to the discussion of $v_{\alpha p}$, one may expect that 
     $v_{\alpha p}$ is largely determined by the proportion of the
     external energy deposited to alpha particles,
     characterized by the parameter $\chi$.
Figure~\ref{fig_vap_par} displays the dependence on $\chi$ of $v_{\alpha p}$
     at several different radial distances as labeled.
Results from 1.5-D models are displayed by thick curves,
      whereas those from 1-D models are given by thin curves for comparison.
Inspection of the solid lines shows that
       the transition from negative to positive
       $v_{\alpha p}$ at 0.3~AU occurs at $\chi=1.6$ in both  models.
With $\chi$ varied between $0.5$ and $3.4$,
       $v_{\alpha p}$ at 0.3~AU goes from 
       $-132$ to $127$\velunits\ for the 1.5-D model,
       while from $-143$ to $134$\velunits\ for the 1-D model.
When comparing the thin and thick solid lines, one can see that
       the profile of $v_{\alpha p}$ for 1-D models differs only slightly
       from that for 1.5-D ones.
Going from 0.3 to 4~AU,  the magnitude of $v_{\alpha p}$ is significantly
       reduced by introducing the azimuthal components.
Unsurprisingly this effect is more prominent at the high and low ends.
For instance the 1-D model for $\chi=3.4$ yields $v_{\alpha p}=120$\velunits\ at 4~AU,
      whereas the corresponding 1.5-D model gives $v_{\alpha p}=21.8$\velunits.
It turns out that introducing the $\phi$ components reduces $v_{\alpha p}$ via
      different mechanisms for different portions of this parameter range.
For $\chi$ between 1.0 and 2.3, the proton-alpha friction eventually
      becomes operative between 0.3 and 4~AU to suppress any significant $|v_{\alpha p}|$.
As a result, no obvious deviation shows up between 1-D and 1.5-D models.
If $\chi \gtrsim 2.3$,  the reduction of $|v_{\alpha p}|$ in the 1.5-D model
      is achieved partly by the enhanced electron pressure gradient force
      (which stems from the enhanced electron temperature due to reduced effective
      electron thermal conductivity in presence of the spiral field)
      and more importantly via the azimuthal force.
On the other hand, for $\chi \lesssim 1$, the reduction in $|v_{\alpha p}|$ is
      solely due to the azimuthal force, since in this case, the electron pressure gradient force
      tends to increase the magnitude of $v_{\alpha p}$ as a consequence of the smaller
      charge-to-mass ratio of alpha particles compared with that of protons.

\subsubsection{Angular Momentum Transport}
\label{sec_num_slow_angmom}

\begin{figure}
\begin{center}
\epsscale{.99}
\plotone{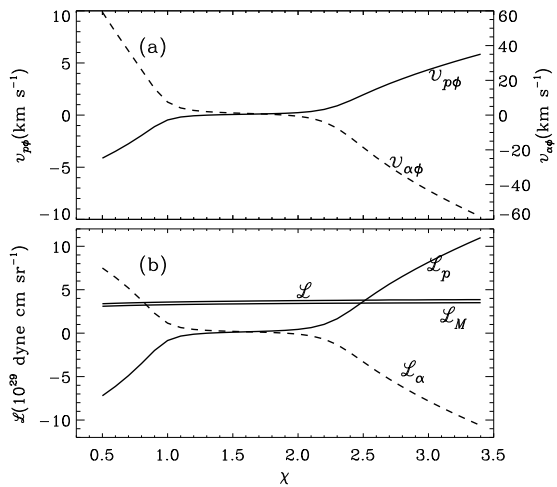}
    \caption{
    Values of several parameters at 1~AU as a function of parameter $\chi$.
    (a) Azimuthal speeds of protons $v_{p\phi}$ and alpha particles $v_{\alpha\phi}$,
    and (b) Specific angular momentum fluxes carried by
        the proton fluid ${\cal L}_p$,
        the alpha fluid ${\cal L}_\alpha$,
        and magnetic stresses ${\cal L}_M$.
	The total flux ${\cal L}$ is also given.
     }
\label{fig_azi_par}
\end{center}
\end{figure}

Figure~\ref{fig_azi_par} displays the distribution with $\chi$ of the values at 1~AU of
      (a) the azimuthal speeds of ions $v_{p\phi}$ and $v_{\alpha \phi}$,
      and (b) specific angular momentum fluxes including that carried by
      protons ${\cal L}_p$, alpha particles ${\cal L}_\alpha$,
      as well as by magnetic stresses ${\cal L}_M$.
From Fig.\ref{fig_azi_par}a one can see that 
      $v_{p\phi}$ and $v_{\alpha\phi}$ at 1~AU tend to have opposite signs, 
      an evidence that they are mainly
      determined by the terms associated with $v_{\alpha p, r}$ in
      Eqs.(\ref{eq_vpphi}) and (\ref{eq_viphi}).
As a matter of fact, for all $\chi$ but $\chi=1.9$, term II in Equation~(\ref{eq_viphi})
      contributes more than term I to $v_{\alpha \phi}$.
As to $v_{p\phi}$, with the exception of $1.4\le \chi \le 2$, the main contribution
      also comes from term II in Equation~(\ref{eq_vpphi}). 
Looking at Fig.\ref{fig_vap_par}, one finds that for this narrow range of $\chi$,
      $v_{\alpha p}$ at 1~AU ranges from -3 to 1.1\velunits.
At substantial relative speeds between protons and alpha particles,
      $v_{p\phi}$ can become as large as $5.84$\velunits\ when $\chi=3.4$,
      for which $v_{\alpha\phi} = -58.5$\velunits.
At the other extreme $\chi=0.5$, $v_{p\phi}$ ($v_{\alpha\phi}$) is -4.1 (59.3)\velunits.
In addition, from Fig.\ref{fig_azi_par}b one can see that 
      ${\cal L}_M$ or ${\cal L}$ varies little,
      although ${\cal L}_p$ and ${\cal L}_\alpha$ vary significantly within the parameter range
      as a consequence of the sensitive dependence of $v_{\alpha\phi}$ and $v_{p\phi}$ on $\chi$.
For instance, when $\chi$ varies from $0.5$ to $3.4$, ${\cal L}$ 
      shows only a slight increase from 3.39 to 
         3.86 $\times 10^{29}$\specangunits,
      and ${\cal L}_M$ increases  from 3.09 to 3.51 $\times 10^{29}$\specangunits.
The increase of ${\cal L}$ with $\chi$ can be explained by Equation~(\ref{eq_calL}).
For the solutions considered, the location of the Alfv\'enic point varies little.
However, the mass loss rate, which is mainly determined by the proton flux,
      increases with increasing $\chi$ as a consequence of more energy being
      deposited to the subsonic portion for the proton gas
      \citep{LeerHolzer_80}.
On the other hand, the ratio ${\cal L}_M/{\cal L}$ shows little variation with $\chi$.
This can be seen as an indication that the ratio of the bulk speed $u_r$ attained asymptotically
      to that at the Alfv\'enic point is rather insensitive to $\chi$,
      since ${\cal L}_M/{\cal L}\approx u_{r, A}/u_r$ in the region $r\gg r_A$
      (cf. Eq.(\ref{eq_distriL2})).
Moreover, if $u_r$ varies little beyond $r_A$, as is the case for all the obtained 
       slow wind solutions, ${\cal L}_M/{\cal L}$ should be only slightly larger than
       unity for $r \gg r_A$.
The expectation is reproduced by Fig.\ref{fig_azi_par}b from which one can see 
       that magnetic stresses are
       always the most important term in ${\cal L}$,
       despite the fact that individual angular momentum fluxes carried by ion species,
       ${\cal L}_p$ or ${\cal L}_\alpha$,
       can be much larger than ${\cal L}_M$.

Now the model results can be compared with the Helios measurements for
      specific angular momentum fluxes
     \citep{Pizzo_etal_83, MarschRichter_84}.
One can see that, although ${\cal L}_M$ falls within the uncertainties
     of the measurements,
     and values like 1.3$\times 10^{29}$\specangunits\ can be found for ${\cal L}_\alpha$
         at $\chi\approx 1$,  
     no values as large as 19.6$\times 10^{29}$\specangunits\
          can be found for ${\cal L}_p$
          unless the alpha particles are heated even more intensely.
Moreover, Fig.\ref{fig_azi_par}b does not reproduce the observed tendency
     for the particle contribution
     ${\cal L}_P$ to be larger than the magnetic one ${\cal L}_M$ at 1~AU.
As has been discussed, the reason is that the solar wind bulk flow $u_r$ experiences little
     acceleration outside the Alfv\'enic point $r_A$.
It is unlikely that this feature can be changed by simply varying other
     model parameters.
Actually the discrepancy between model predictions and measurements on the relative importance
     of ${\cal L}_P$ and ${\cal L}_M$ is common to all existing models that treat the meridional
     and azimuthal dynamics self-consistently
     (cf. sections I and VII of \citet{Pizzo_etal_83}).
{To examine the possible sources leading to this discrepancy, in what follows we consider
     some of the physics that is not accounted for in the present model.}

{In the framework of steady state models, the fast and slow magnetosonic waves,
      in the WKB approximation, 
      have non-zero $r\phi$
      components in their stress tensors, thereby also contributing to the angular momentum loss
      of the solar wind \citep{Marsch_86}.
On the one hand, this contribution may be comparable to that due to the
     background
     flow and magnetic field stress when the waves have sufficiently large
     amplitudes.
On the other hand, depending on the propagation angle with respect to the background
     magnetic field, the particle part may dominate
     the magnetic one in the wave contribution.
(This can be seen if one evaluates the $r\phi$ component of Eq.(7)
      by using, e.g., the fast wave eigen-function Eq.(10)
      in \citet{Marsch_86}).
Therefore, including compressional waves may provide a possible means to resolve the
     apparent discrepancy.
Furthermore, extending the WKB analysis to higher order, \citet{Hollweg_73} found that 
     Alfv\'en waves with finite wavelength may carry a non-vanishing flux of angular
     momentum, but the contribution seems to worsen rather than improve
     the discrepancy.
Finally, we} refer the reader to the discussion of \citet{Hu_etal_03b}, whose
     2.5-D model also incorporates the momentum deposition by Alfv\'en waves.
Instead of attributing this discrepancy to mechanisms {missing} in the steady state
     model, \citet{Hu_etal_03b} proposed that the tendency for ${\cal L}_P$ to be larger than
     ${\cal L}_M$ is due to
     interplanetary dynamical processes such as the forward fast shocks traversing the solar wind.
Their discussion also applies to the present model, although {the details 
     of the response of the three-fluid plasma to fast shocks need to be explored}.

\section{SUMMARY}
\label{sec_conc}
The near-ecliptic measurements of the fast solar wind by Helios demonstrated
        the existence of a significant proton-alpha differential speed, $v_{\alpha p}$,
        varying from 150\velunits\ at 0.3 AU to 30-40\velunits\ at 1 AU\citep{Marsch_etal_82}. 
A steady decrease of $v_{\alpha p}$ between 1.5 and 4 AU was also measured by
        Ulysses out of the ecliptic plane \citep{Reisenfeld_etal_01}. 
The Helios measurements also indicated the existence of two categories of slow solar wind. 
In one category, the alpha particles tended to be slower than the protons, whereas
       the opposite trend was found in the other category\citep{Marsch_etal_82}.
The measurements also yielded information about the angular momentum transport, 
       in particular, the specific angular momentum fluxes in the magnetic stress ${\cal L}_M$,
       and in the two major ion species
       ${\cal L}_p$ and ${\cal L}_\alpha$\citep{Pizzo_etal_83, MarschRichter_84}.

In an attempt to account for these observations, we introduced solar rotation self-consistently
      in a three-fluid solar wind model. 
The impetus for this approach was inspired by the fact that {the solar rotation} is an inherent process, 
      independent of different mechanisms invoked for accelerating the solar wind. 
By exploring the resulting interplay between $v_{\alpha p}$ and the angular momentum transport
      we reached the following conclusions:
\begin{enumerate}
 \item 
The force introduced in the radial momentum equations by the azimuthal components can play a
        significant role in the force balance in interplanetary space. 
Its main effect is to reduce the difference between the ion flow speeds.
\item 
While the effect of solar rotation could not account for the decrease in
        $v_{\alpha p}$ in the fast solar wind from 0.3 to 1~AU, it could account for
        the $v_{\alpha p}$ profile measured by Ulysses beyond 2~AU if an appropriate
        value for $v_{\alpha p}$ was chosen at 2~AU.
\item 
The effect of solar rotation was found to be more pronounced for the slow solar wind
       if a significant $v_{\alpha p}$ developed by 0.3 AU, 
       resulting in a relative change of 10-16\% in the radial speed of the alpha particles
       between 1 and 4 AU.
\item 
For the slow solar wind solutions considered, the angular momentum flux carried by
       alpha particles, ${\cal L}_\alpha$, at 1 AU, was almost exclusively determined by $v_{\alpha p}$.
       Moreover, ${\cal L}_p$ was also determined by $v_{\alpha p}$ for
       $v_{\alpha p} \gtrsim 1$ or $\lesssim -3$\velunits.
\item 
In the fast solar wind, ${\cal L}_p$ and ${\cal L}_\alpha$ beyond 0.3~AU are also mainly
       due to $v_{\alpha p}$. 
This suggests that even though alpha particles take up only a small fraction of the energy
       and linear momentum fluxes of protons, they cannot be neglected when the proton angular
       momentum flux is concerned, i.e. ${\cal L}_p$ cannot be studied by simply incorporating
       the azimuthal components into models of an electron-proton solar wind. 
\item 
For the slow solar wind, ${\cal L}_p$ and ${\cal L}_\alpha$ can be several times larger
       in magnitude than the flux carried by the magnetic stresses ${\cal L}_M$. 
However, the total particle distribution ${\cal L}_P= {\cal L}_p +{\cal L}_\alpha$
       is always smaller than ${\cal L}_M$ for all solutions. 
While this tendency is inherent to the present model, it is at variance
       with the Helios measurements.
\end{enumerate}

{In closing, some words of caution seem in order. 
As has been discussed, a pure fluid model of a quiet solar wind as adopted in this study
      is rather idealized.
In reality, the solar wind will be subject to various disturbances such as
      compressional magnetosonic waves, shock waves, and 
      other processes resulting from stream-stream interactions.
These disturbances may significantly modify the distribution of 
      the angular momentum flux between particles and magnetic field stresses,
      and may alter the overall angular momentum loss rate of the solar wind.
With increasing distance, such as in the outer heliosphere, 
      these disturbances are expected to become increasingly important.
With this caveat in mind, the presented study nevertheless provides
      a better understanding of the underlying physics.
Even when the above-mentioned dynamical processes are taken into account, one would expect to see that 
      solar rotation has significant effects
      on the alpha flow speeds in interplanetary space, and in turn, 
      alpha particles are important in the problem of angular momentum transport
      of the solar wind.
}

\acknowledgements
This work is supported by a PPARC rolling grant to the University of Wales Aberystwyth.
{We wish to thank the anonymous referee for the constructive comments that helped 
    improve the manuscript.}

\end{document}